\RequirePackage{fix-cm}
\documentclass[smallextended]{svjour3}       
\smartqed  
\usepackage{amsmath, bm}
\usepackage{amssymb}
\usepackage{graphicx,psfrag,epsf}
\usepackage{enumerate}
\usepackage{natbib}
\usepackage{url} 

\usepackage{graphicx}

\usepackage{hyperref}
\usepackage{float}
\usepackage{subfig}
\usepackage{wrapfig}
\usepackage{rotating}
\usepackage{lscape}
\usepackage[T1]{fontenc}

\usepackage{amssymb}
\usepackage{tabularx}
\usepackage[dvips]{epsfig}

\usepackage{dsfont}

\usepackage{slashbox}
\usepackage{multirow}
\usepackage{xcolor}
\usepackage{lineno}      

\usepackage{lscape}
\usepackage{amssymb}
\usepackage{tabularx}
\usepackage[dvips]{epsfig}

\usepackage{graphicx}
\usepackage{dsfont}
\usepackage{url}
\usepackage{slashbox}
\usepackage{multirow}
\usepackage{verbatim}

\usepackage[symbol]{footmisc}

\newcommand*{\rom}[1]{\expandafter\@slowromancap\romannumeral #1@}

    \usepackage{hyperref} 
\hypersetup{colorlinks=true, linkcolor=blue,
  citecolor=blue,urlcolor=blue} 

\usepackage{doi}

\newcommand{\vertiii}[1]{{\left\vert\kern-0.25ex\left\vert #1 
    \right\vert\kern-0.25ex\right\vert_F}}

\renewcommand{\thefootnote}{\arabic{footnote}}

\newcommand{\Rhat}{\widehat{R}_{j_1, j_2}}

\newcommand{\PP}{\mathbb{P}}

\newcommand{\EE}{\mathbb{E}}

\usepackage{color} 
\usepackage[normalem]{ulem} 

    \newcommand{\tfsib}[1]{#1} 
    \newcommand{\ta}[1]{#1} 
    \newcommand{\td}[1]{} 

\begin{document}

\title{EPISTASIS DETECTION VIA THE JOINT CUMULANT
}


\author{
    Randall Reese    \and
    Guifang Fu  \footnotemark[1] \and
    Geran Zhao \and
	  Xiaotian Dai \and
	  Xiaotian Li \and
	  Kenneth Chiu 
	  }


\institute{Randall Reese \at
              Idaho National Laboratory,\\
              Idaho Falls, Idaho 83401, USA
           \and
           Guifang Fu \at
              Department of Mathematical Sciences,\\
	            SUNY Binghamton University,\\
    	        Binghamton, NY 13902, USA
	         \and
           Geran Zhao \at
              Department of Mathematical Sciences,\\
	            SUNY Binghamton University,\\
    	        Binghamton, NY 13902, USA
	         \and
           Xiaotian Dai \at
              Department of Mathematical Sciences,\\
	            SUNY Binghamton University,\\
    	        Binghamton, NY 13902, USA
    	     \and
    	     Xiaotian Li \at
              Department of Computer Science,\\
	            SUNY Binghamton University,\\
    	        Binghamton, NY 13902, USA
	        	     \and
    	    Kenneth Chiu \at
              Department of Computer Science,\\
	            SUNY Binghamton University,\\
    	        Binghamton, NY 13902, USA
}

\date{Received: date / Accepted: date}

\renewcommand{\thefootnote}{\fnsymbol{footnote}}
\footnotetext[1]{Correspondence to Guifang Fu: gfu@binghamton.edu}

\maketitle

\begin{abstract}
Selecting influential nonlinear interactive features from ultrahigh dimensional data has been an important task in various fields. However, statistical accuracy and computational feasibility are the two biggest concerns when more than half a million features are collected in practice. Many extant feature screening approaches are either focused on only main effects or heavily rely on heredity structure, hence rendering them ineffective in a scenario presenting strong interactive but weak main effects. In this article, we propose a new interaction screening procedure based on joint cumulant (named JCI-SIS). We show that the proposed procedure has strong sure screening consistency and is theoretically sound to support its performance. Simulation studies designed for both continuous and categorical predictors are performed to demonstrate the versatility and practicability of our JCI-SIS method. We further illustrate the power of JCI-SIS by applying it to screen \tfsib{27,554,602,881 interaction pairs involving 234,754 single nucleotide polymorphisms (SNPs)} for each of the 4,000 subjects collected from polycystic ovary syndrome (PCOS) patients and healthy controls.

\end{abstract}

\section{INTRODUCTION}
\ta{With the rapid advances in high-throughput data collection techniques}, there has arisen an accompanying need to \ta{select truly influential features from an extremely large-scale candidate pool \citep{Liu2015selective, chernoff2009discovering}.  \cite{FanLv:2008} pioneered the sure independence screening (SIS) approach, following which an abundance of powerful feature screening approaches possessing the theoretical sure screening properties have been developed \citep{FanFanHiDim:2008, Wang2009forward, FanFengSong:2011, zhuLiLiZhu2011, Kim2012consistent, SuperLongNames, XuChen:2014, DCSIScitation, MMLE2010, Huang2014, ni2016entropy, pan2019generic}. However, as stated by \cite{HaoZhangiFORM}, the majority of current feature screening literature focuses on selecting individual features with only main effects.}

\ta{Since relationship between predictors and the response is often more complicated than can be captured by additive main effects alone, developing techniques to consider interactions between predictors is crucial. In particular, the importance of two-way interactions has been recognized not only in statistics literature \citep{HaoZhangiFORM, Fan2015innovated}, but also in various applied fields \citep[See e.g.][]{Phillips2008epistasis, ueki2012ultrahigh, Li2014fast, Wei2014detecting, Karkkainen2015efficient, Gosik2017iform, Fang2017tsgsis}. Nevertheless, two challenges deter interaction screening from being well studied.} 

\ta{The first challenge is the unusual computational cost when screening important interactions from ultrahigh dimensional data  \citep{FanLi:BigData2006, FanHanLiu:BigData}. \tfsib{Specifically, an exhaustive search of all possible interactions requires examining $\binom{p}{2}$ potential pairs, which incurs an astronomic level of computational cost when the number of features (denoted as $p$) is ultrahigh dimensional} \citep{HaoZhangiFORM, Fan2015innovated,Fang2017tsgsis,Hao2018model}.} For example, nearly 50 million pairs are involved for a relatively small value of $p = 10,000$. When $p = 45,000$, the number of interactions to be considered numbers in the trillions. \ta{The second challenge comes from lack of applicable approaches. Many existing classic approaches designed for interaction may lose power or become increasingly unstable as the dimension dramatically grows \citep{Fan2015innovated}. For example, interaction selection approaches such as joint analysis examined all marginal and interaction effects in a single global search \citep{Yuan2007efficient, Yuan2009structured, Zhao2009composite, Choi2010variable}, which is only feasible when the feature space is not high dimensional \citep{HaoZhangiFORM}.} 
\cite{Sariyar2014combining} examined the use of random forests in screening time-to-event data for interactions. They claimed that random forests works well when the feature space is medium dimensional, but as the feature space becomes ultrahigh dimensional, it may become unstable or computationally intractable \citep{Ziegler2007data, Zhang2009willows, Schwarz2010safari, Huang2014}. 
 
\ta{To overcome these aforementioned challenges of ultrahigh dimensional data analyses in detecting important two-way interaction pairs, many approaches first extract important main effects and then examine only a subset of interactions having at least one influential main effect \citep{Li2014fast, Fang2017tsgsis, HaoZhangiFORM, Hao2018model}. While these approaches significantly reduce the number of interactions that are to be examined, they pose excess reliance on an \emph{a priori} heredity structure. Specifically, a \emph{weak heredity} means that $(X_{j_1},X_{j_2})$ is considered only if at least one of the main effects, $X_{j_1}$ or $X_{j_2}$, is individually associated with the response. A \emph{strong heredity} requires both main effects are influential. As such, any method that relies on a heredity structure will neglect a scenario where there exists a strong interactive effect, but for which both corresponding main effects are very weak \citep{Marchini2005genome, ueki2012ultrahigh, RITCHIEmdr2001}. However, handling such instances is found to be important in many practical applications \citep{Balding2006tutorial, Marchini2007new, Cordell2009detecting, manolio2009, ueki2012ultrahigh, Fang2017tsgsis}.  Therefore, the ideal interaction screening method should not \emph{a priori} assume a weak or strong heredity structure.} 

\ta{In this article, we propose a novel feature screening approach based on joint cumulant (JCI-SIS for short) to select influential interactions from an ultrahigh dimensional candidate pool (defined as $\log(p) = \mathcal{O}(n^\xi)$ for some constant $\xi>0$, where $n$ is the sample size). JCI-SIS not only overcomes the two challenges as outlined above but also completely free of the restrictions of heredity structure. We prove that the JCI-SIS satisfies the strong screening consistency under a set of reasonable conditions, which guarantees that the subset selected by the JCI-SIS accurately captures the truth when sample size is large enough. We also empirically demonstrate that JCI-SIS is computationally effective for an extremely large real dataset with \tfsib{27,554,602,881 interaction pairs involving 234,754 single nucleotide polymorphisms (SNPs) for each of the 4,000 subjects collected from polycystic ovary syndrome (PCOS) patients and healthy controls}.} 

\ta{JCI-SIS is compared with two other approaches existing in current feature screening literature. These methods are most relevant in that they also aim for selecting interactions from ultrahigh dimensional data and possess sure independence screening consistency. Specifically, the iterative forward selection method (iFORM) proposed by \cite{HaoZhangiFORM} recognizes interactions based on important main effects using stepwise forward selection and Bayesian information criterion (BIC) criteria.  The generalized Pearson correlation approach (GPC-SIS in short) given by \cite{Huang2014} extends the marginal Pearson correlation from one predictor to a pair. These two approaches both rely on the assumption of strong heredity and also are restricted to certain data types (iFORM for continuous data and GPC-SIS for categorical data). We empirically demonstrate that JCI-SIS dramatically outperforms iFORM and GPC-SIS in terms of accuracy and speed through finite sample simulation studies.} 

The remainder of this paper is outlined as follows: \ta{In Section \ref{prelims} we detail our proposed JCI-SIS approach}, including statements on its theoretical properties. This is followed by an assessment in Section \ref{simulations} of the finite sample performance of JCI-SIS via multiple simulation studies and a real data analysis. In Section \ref{discuss} we provide concluding remarks on our method and our findings. Finally, \ta{the Appendix} contains the proofs of the theorems presented in Section \ref{prelims}.

\section{\ta{STRONG INTERACTION SCREENING USING JOINT CUMULANT}}\label{prelims}
\subsection{Some Preliminaries}

\ta{Let $(Z_1, Z_2, \ldots, Z_r)$ be a $r$-dimensional random vector. The $r$-way joint cumulant  of $(Z_1, Z_2, \ldots, Z_r)$ is defined as \citep{SLJHuCum1991, NicaCum2006}
\[
\kappa_r(Z_1, Z_2, \ldots, Z_r) = \sum_{\pi} (|\pi|-1)! (-1)^{|\pi|-1} \prod_{B \in \pi} E(\prod_{i \in B} Z_i),
\]
where $\pi$ runs through the list of all partitions of $\{1,\ldots,r\}$, $B$ runs through the list of all blocks of the partition $\pi$, and $|\pi|$ is the number of parts in the partition. 
}

\ta{Specific to our purposes here, the three-way joint cumulant, written as $\kappa_3(\cdot, \cdot, \cdot)$, for a $3$-dimensional random vector is given as follows:
\begin{eqnarray*}\kappa_3(Z_1, Z_2, Z_3) &=& \EE(Z_1Z_2Z_3) - \EE(Z_1Z_2)\EE Z_3-\EE(Z_1Z_3)\EE Z_2\\
&& \quad \quad \quad \quad \quad ~-\EE(Z_2Z_3)\EE Z_1 + 2 \EE Z_1\EE Z_2\EE Z_3.\end{eqnarray*}} Note that $\kappa_3(Z_1, Z_2, Z_3)$ is zero if any one of the variables is statistically independent from the other two.

\ta{Similarly, the two-way joint cumulant, written as $\kappa_2(\cdot, \cdot)$, for two random variables is defined as
\[
\kappa_2(Z_1, Z_2) = \EE\left[(Z_1 - \EE Z_1)(Z_2 - \EE Z_2)\right],
\]
}
which is \ta{actually} the covariance between two random variables (\ta{i.e.,} the variance when both arguments are equal).

\ta{\subsection{An Interaction Ranking and Screening Procedure}
In this section we propose a new interaction screening procedure. Let $Y$ be the response variable and $\{X_j; j=1,\ldots, p\}$ be the $j^{th}$ predictor, where $p$ is the total number of predictors. 
}

\ta{Define the screening criterion $R_{j_1, j_2}$}, admitting three random variables as arguments, as
\begin{equation}
R_{j_1, j_2} = \frac{|\kappa_3(Y, X_{j_1}, X_{j_2})|}{\sqrt{\kappa_2(X_{j_1},X_{j_1} )\kappa_2(X_{j_2}, X_{j_2})\kappa_2(Y,Y)}},~~j_1<j_2;~j_1,j_2=1,\ldots,p.
\label{Rjj}
\end{equation}
\ta{We are motivated to utilize $R_{j_1, j_2}$ as a feature screening criterion because $R_{j_1, j_2}$ is zero if 
 the response $Y$ is independent of the predictor pair $(X_{j_1},X_{j_2})$.}

We estimate $R_{j_1, j_2}$ as
\begin{equation}
\widehat{R}_{j_1, j_2}=\frac{\sqrt{n}\left|\sum\limits_{i = 1}^n(X_{ij_1} - \overline{X}_{j_1})(X_{ij_2} - \overline{X}_{j_2})(Y_i - \overline{Y})\right|}{\sqrt{\left(\sum\limits_{i = 1}^{n}(X_{ij_1} - \overline{X}_{j_1})^2\right)\left(\sum\limits_{i = 1}^{n}(X_{ij_2} - \overline{X}_{j_2})^2\right)\left(\sum\limits_{i = 1}^n(Y_i-\overline{Y})^2\right) }},
\label{Rhat}
\end{equation}
\ta{where $X_{ij_1}$, $X_{ij_2}$, and $Y_i$ are sample observations and $\overline{X}_{j_1}$, $\overline{X}_{j_2}$, and $\overline{Y}$ are sample means of $X_{j_1}$, $X_{j_2}$, and $Y$, respectively.}

\ta{Let $\mathcal{S}_F = \{(j_1,j_2)~|~j_1< j_2;~j_1,j_2=1,\ldots,p\}$ denote the indices of the full model, which refers to the unique pairs of all of the predictors except the same predictor with itself. Let $\mathcal{S} \subseteq \mathcal{S}_F$ denote the indices of an arbitrary interaction model under consideration, which is a subset of $\mathcal{S}_F $. Let $\bm{X}_{(\mathcal{S})}$ be the corresponding interaction model or set of predictor pairs that are referred by $\mathcal{S}$.}

Pairs of predictors are then ordered \ta{based on the association strength score} $\widehat{R}_{j_1, j_2}$. The interaction pair with the largest $\widehat{R}_{j_1, j_2}$ has \ta{the strongest association strength} with the response $Y$ and \ta{the pairs with smallest scores are not important to the response}. \ta{Since the absolute value is used, $\widehat{R}_{j_1, j_2}$ is non-negative. The output of the JCI-SIS feature screening procedure is the selected indices of the subset given by 
\begin{equation}
\widehat{\mathcal{S}} = \{(j_1,j_2)| j_1<j_2;~j_1,j_2=1,\ldots,p;~ \widehat{R}_{j_1, j_2} > c\},
\label{Shat}
\end{equation}
where $c$ is a pre-specified threshold value. 
} 

Let $\mathcal{D}\left(Y \mid \bm{X}_{(\mathcal{S})}\right)$ \ta{denote} the conditional distribution of $Y$ given \ta{a set of predictor pairs} $\bm{X}_{(\mathcal{S})}$. A model $\bm{X}_{(\mathcal{S})}$ is sufficient if
\[\mathcal{D}\left(Y \mid \bm{X}_{(\mathcal{S}_F)}\right) = \mathcal{D}\left(Y\mid\bm{X}_{(\mathcal{S})}\right).\]
The full model $\mathcal{S}_F$ is trivially sufficient. Our aim \ta{in feature screening} is to determine the \textit{smallest} model that contains the interaction features truly influential to the response. In the following we will denote the true model by $\mathcal{S}_T$ and an estimated model \ta{output from JCI-SIS} by $\widehat{\mathcal{S}}$.

\subsection{Theoretical Properties}\label{thrtProp}
\ta{The theoretical properties of the proposed independence screening procedure JCI-SIS will be claimed in this section.} We establish four conditions that will aid us in \ta{the technical proofs.}
\begin{enumerate}
    \item[(C1)] \emph{Lower bound on the standard deviations}. We assume that there exists a positive constant $\sigma_{\text{min}}$ such that \ta{for all $j_1,j_2=1,\ldots,p$,
   \[\min\Bigg\{\sqrt{\kappa_2(X_{j_1}, X_{j_1})}, \sqrt{\kappa_2(X_{j_2}, X_{j_2})},\sqrt{\kappa_2(Y, Y)}\Bigg\} > \sigma_{\text{min}}>0,\]}
    \ta{which} excludes \ta{constant} features that have a standard deviation of 0. \ta{Condition (C1) is crucial to guarantee that the definition of $R_{j_1, j_2}$  defined in Equation (\ref{Rjj}) is valid.}
   \item[(C2)] \emph{Upper bound on the standard deviations}. We assume that
   \ta{ \[\max\Bigg\{\sqrt{\kappa_2(X_{j_1}, X_{j_1})}, \sqrt{\kappa_2(X_{j_2}, X_{j_2})},\sqrt{\kappa_2(Y, Y)}\Bigg\} < \sigma_{\text{max}} < \infty,\] for all $j_1,j_2=1,\ldots,p$.} 
\tfsib{ \item[(C3)] \textit{\ta{Lower bound on the} joint cumulant association}. We assume that there exists some positive constants $\gamma >0$ and $\zeta > 0$ such that \[| \kappa_3(X_{j_1}, X_{j_2}, Y) | > \gamma n^{-\zeta},\] whenever
$(j_1, j_2) \in S_T$.  For such choices of $\gamma$ and $\zeta$, let $\delta_{\min} = \gamma n^{-\zeta} $.  That is to say, there exists some positive constant $\delta_{\min}$ such that \[\min_{(j_1, j_2) \in S_T} | \kappa_3(X_{j_1}, X_{j_2}, Y) |  > \delta_{\min}>0.\] }
  
    \item[(C4)] \textit{Existence}. \ta{We assume that $R_{j_1, j_2} < \infty$ for all $\{(j_1,j_2)| j_1<j_2;~j_1,j_2=1,\ldots,p\}$. In addition, we assume that $R_{j_1, j_2} = 0$ for any pair of indices $(j_1, j_2) \not\in \mathcal{S}_T$.}
 \end{enumerate}

\ta{Once these four conditions are satisfied, the following theorems will be established to guarantee the \emph{strong screening property} for the JCI-SIS procedure.}

\begin{theorem}\label{Thm2} (\textit{Sure Screening Consistency}). Given  that conditions (C1), (C2), and (C3) still hold, while removing from (C4) only the assumption of $R_{j_1, j_2} = 0$ for all $(j_1, j_2) \notin \mathcal{S}_T$, there exists a positive constant $c>0$ such that \[\mathbb{P}(\mathcal{S}_T  \subseteq \widehat{\mathcal{S}}) \longrightarrow 1 \text{ as } n \longrightarrow \infty.\]
(But $\mathbb{P}(\widehat{\mathcal{S}} \subseteq \mathcal{S}_T)$ may not converge to 1 as $n$ approaches infinity).
\end{theorem}
\begin{theorem}\label{Thm1} (\textit{Strong Sure Screening Consistency}). Given conditions (C1), (C2), (C3) and (C4), there exists a positive constant $c>0$ such that \[\mathbb{P}(\widehat{\mathcal{S}} = \mathcal{S}_T) \longrightarrow 1 \text{ as } n \longrightarrow \infty.\]
\end{theorem}

\ta{Most of the feature screening approaches only claim the sure screening property because it is adequate to guarantee that the truth is not missed. However, the strong sure screening consistency (harder to prove than the sure screening property) not only ensures that the true model is not missed but also guarantees that the selected subset is the smallest.} The proofs of these two theorems are presented in the Appendix.

\ta{As a byproduct of the proofs of Theorems 1 and 2, we also claim two corollaries. Although they do not directly deal with sure screening, they allow us to investigate more theoretical properties of the proposed JCI-SIS method.}

\begin{corollary}\label{Cor1}
There exists a value $R_{\min}$ such that for any pair $(j_1, j_2) \in \mathcal{S}_T$, we have $R_{j_1, j_2} > R_{\min}$. 
\end{corollary}
\ta{This will be shown in Step 1 of the proofs of Theorems 1 and 2.}

\begin{corollary}\label{Cor2}
$\Rhat$ converges \textit{uniformly} in probability  to $R_{j_1, j_2}$. In other words,
\[\mathbb{P}\left(\max_{(j_1, j_2)}|\Rhat - R_{j_1, j_2}| > \varepsilon\right) \rightarrow 0 \quad \text{ as } n\rightarrow\infty\] for any $\varepsilon> 0$ and for \ta{all $1 \leq j_1<j_2 \leq p$. 
\end{corollary}
This will be shown in Step 2 in the proofs of Theorems 1 and 2.}

\section{SIMULATION STUDIES AND EMPIRICAL DATA ANALYSIS}\label{simulations}
\ta{In this section we perform four Monte Carlo simulation studies under various scenarios to empirically validate the performance of our proposed JCI-SIS approach in both categorical and continuous data types. Specifically, we  accurately demonstrate the capability of JCI-SIS in handling the most challenging scenario, namely that in which strong interactive effects exist, but neither involved predictor exhibits a relevant main effect (Simulation studies 1 and 2). We also test the performance of JCI-SIS when main effects can be counted on to detect interactions, which is the heredity structure that current existing interaction feature screening approaches assume (Simulation studies 3 and 4). We find that, even given a hereditary structure (with which existing methods should ostensibly excel), JCI-SIS still performs well.

In Simulation studies 1 and 3, we consider categorical data and compare JCI-SIS with GPC-SIS. The performance is assessed through the average and median ranking of each true interaction pair across all simulation replications. The closer to the true model, the better performance the screening procedure has. In Simulation studies 2 and 4, we consider continuous data and compare JCI-SIS with iFORM. To compare the performance of both methods, we report the percentage of replicates for which a given method found the true interaction pairs to be among the top five most important interactions. The closer to 100\% the
reported percentage is, the more accurate the method can be said to be.

For the first three simulation studies, we keep the sample size to be 200 and the number of individual predictors $p$ to be 1,000 (corresponding to 499,500 pairs). For the last simulation study, we set $n=100$ and $p=500$ (corresponding to 124,750 pairs). We replicate each simulation study 100 times.} Each of these simulation studies, as well as the associated results, are summarized below. 

\subsection{Simulation Study 1}
We generate each predictor $X_j$ \ta{by a Bernoulli process}, with each outcome being equally likely. \ta{We design the first two predictors to be truly influential to the response $Y$, i.e., $\mathcal{S}_T=\{(1,2)\}$} and
\[Y = X_1 \ast X_2.\] 
 The results for the Simulation Study 1 are summarized in Table \ref{table:Sim1} below.

\begin{table}[h!]
\caption{The estimated mean and median rank of $(X_1, X_2)$ in the Simulation Study 1.}\label{table:Sim1}
\begin{center}
\begin{tabular}{ |c | c c | }
\hline
 & JCI-SIS & GPC-SIS  \\ \hline\hline
Mean Rank of $(X_1, X_2)$ & 1 & 2104.5 \\
Median Rank of $(X_1, X_2)$ &1 &1306\\
  \hline
\end{tabular}
\end{center}
\end{table}
Note that \ta{JCI-SIS perfectly locates the true interaction pair $X_1 \ast X_2$ as the top one among all simulation replicates. This means that JCI-SIS perfectly detects the one true interactive pair in each of the 100 replications. On the contrary, GPC-SIS fails prodigiously by ranking thousands of other predictor pairs prior to the truth}. Both the average and the median rankings of $X_1 \ast X_2$ got by GPC-SIS are much too large for GPC-SIS to be considered a reliable feature screening approach in this case.

\subsection{Simulation Study 2}
The Simulation Study 2 is very similar in form to the Simulation Study 1 \ta{but has two differences. To verify the power of JCI-SIS with continuous variables and multiple interactions, we generate $X_j$ independently from normal distribution as} $X_j \overset{\text{i.i.d.}}{\sim} N(\mu = 0, \sigma = 2), j=1,\ldots,p.$ \ta{In addition, we  connect $Y$ with two pairs of interactions by} \[Y = X_1 \ast X_2 + X_3 \ast X_4.\] 

\ta{The results of this Simulation Study show the remarkable difference between JCI-SIS and iFORM in determining true interactive effects when no main effects are prevalent (see Table \ref{table:Sim3}). iFORM is unable to rank any of the true interactions within top five, but JCI-SIS successfully acquires all of them with 100\% accuracy. }

\begin{table}[h!]
\caption{The percentage of replicates in the Simulation Study 2 for which the given true interaction(s) is found to be one of the five most important interactions.}\label{table:Sim3}
\begin{center}
\begin{tabular}{ |c | c c | }
\hline
 & JCI-SIS & iFORM  \\ \hline\hline
$(X_1, X_2)$ &100\% &0\%  \\
$(X_3, X_4)$ &100\% & 0\% \\
$(X_1, X_2)$ \& $(X_3, X_4)$ & 100\% & 0\%\\
  \hline
\end{tabular}
\end{center}
\end{table}

\subsection{Simulation Study 3} \ta{In an attempt to make a fair comparison between JCI-SIS and GPC-SIS, we closely imitates a simulation study used to test GPC-SIS in Huang et al. (2014). This will allow us to test the performance of JCI-SIS in a scenario where both main and interactive effects exist. We generate the response $Y_i$ and predictors from a binary distribution. Specifically, $Y_i$  is generated from $\PP(Y_i = 1) = 0.75$, and then $X_{ij} \in \{0,1\}$ for $j = 1,3,5,7$ are generated based on $Y_i$} as follows:
\begin{itemize}
\item[$\bullet$] Conditional on $Y_i = k$, \ta{$X_{ij},~j=1,3,5,7$ is generated from $\PP(X_{ij} = 1 | Y_i = k) = \theta_{kj}$, where $\theta_{kj}$ is given in Table \ref{table:Sim2Theta}.}

\begin{table}[h]
\caption{The $\theta_{kj}$ values for generating data in the Simulation Study 3.}\label{table:Sim2Theta}
\centering
\begin{tabular}{|l| cccc|}
\hline
 & \multicolumn{4}{c|}{j} \\ \cline{2-5}
$\theta_{kj}$ & 1 & 3 & 5 & 7 \\ \hline\hline
$k = 0$ & 0.3 & 0.4 & 0.5 & 0.3 \\
$k = 1$ & 0.95 & 0.9 & 0.9 & 0.95 \\ \hline
\end{tabular}
\end{table}

\item[$\bullet$] Given $Y_i$ and $X_{i, 2m-1}$ (for $m = 1,2,3,4$), we generate $X_{i,2m}\in \{0,1\}$ using the following probabilities:
\small
\[\PP(X_{i, 2m} = 1 | Y_i =k, X_{i, 2m-1} = 0) = 0.6I(\theta_{k,2m-1} > 0.5) + 0.4I(\theta_{k,2m-1} \leq 0.5);\]
\[\PP(X_{i, 2m} = 1 | Y_i =k, X_{i, 2m-1} = 1) = 0.95I(\theta_{k,2m-1} > 0.5) + 0.05I(\theta_{k,2m-1} \leq 0.5),\]
\normalsize where $I(\cdot)$ is the standard indicator function.

\item[$\bullet$]  We randomly sample all remaining \ta{non-influential} predictors (i.e. $X_j$ for $j > 8$), \ta{with equally likely outcome}.
\end{itemize}
\ta{Under this design, the influential factors are $X_1$, $X_3$, $X_5$, $X_7$, $(X_1, X_2)$, $(X_3, X_4)$, $(X_5, X_6)$, and $(X_7, X_8)$.}

\begin{table}[h!]
\caption{The mean and median ranking of the true interactions in the Simulation Study 3.}\label{table:Sim2}
\begin{center}
\begin{tabular}{ |c | c c | }
\hline	
 & JCI-SIS & GPC-SIS  \\ \hline\hline
Mean Rank of $(X_1, X_2)$ & 2.01 & 7302.73 \\
Median Rank of $(X_1, X_2)$ &2 &534\\\hline
Mean Rank of $(X_3, X_4)$ & 3.53 & 2365.05 \\
Median Rank of $(X_3, X_4)$ &3 &42.5\\\hline
Mean Rank of $(X_5, X_6)$ & 4.65 &936.65 \\
Median Rank of $(X_5, X_6)$ &4 &16.5\\\hline
Mean Rank of $(X_7, X_8)$ & 2.33& 6563.83 \\
Median Rank of $(X_7, X_8)$ &2 &1083.5\\
  \hline
\end{tabular}
\end{center}
\end{table}
\ta{The results of the Simulation Study 3 are given in Table \ref{table:Sim2}. Since it is impossible for every truly influential interaction to be consistently ranked as the absolute top interaction, any method ranking each true interaction on average in the top five interactions can easily be said to be producing accurate results. In this regard, JCI-SIS obtains excellent results. Contrarily, the average ranking of each influential interaction by GPC-SIS does not lend to confidence in being able to select the true interactions in many cases. Although the median rank of each interaction by GPC-SIS is better than its average respective rank, the reliability and stability of the method is, on the whole, questionable. The large difference between the mean and median rankings obtained from the GPC-SIS indicates its instability over the course of 100 replications. This indicates that GPC-SIS lacks the necessary robustness as an interaction feature screening method.} \tfsib{The run time is 920 seconds for JCI-SIS and 92746 seconds for GPC-SIS for the 100 replications of the Simulation Study 3.}

\subsection{Simulation Study 4}
\ta{Similar to the Simulation Study 3, we continue to verify the power of JCI-SIS in screening interactive features in the presence of individual main effects. However, instead of the categorical data of the Simulation Study 3, we now examine the performance of JCI-SIS on continuous data.} We \ta{generate 100 samples for 500 predictors following} the multivariate normal distribution with mean vector \textbf{0} and $\text{cov}(X_{j_1}, X_{j_2}) = 0.1^{|j_1 - j_2|}$ for $1 \leq j_1, j_2 \leq 500$. \ta{We then set two interactive and four main effects to be influential:} \[Y = X_1 + X_3 + X_6 + X_{10} + 3~X_1 \ast X_3 + 3~X_6 \ast X_{10}.\]

 \begin{table}[t!]
\caption{The percentage of replicates in the Simulation Study 4 for which the given true interaction(s) is found to be one of the five most important interactions.}\label{table:Sim4}
\begin{center}
\begin{tabular}{ |c | c c | }
\hline
 & JCI-SIS & iFORM  \\ \hline\hline
$(X_1, X_3)$ &92\% &21\%  \\
$(X_6, X_{10})$ &92\% & 19\% \\
$(X_1, X_3)$ \& $(X_6, X_{10})$ & 84\% & 9\%\\
  \hline
\end{tabular}
\end{center}
\end{table}

 The results of the Simulation Study 4 are given in Table \ref{table:Sim4}. \ta{As we can see,} iFORM could not accurately and consistently detect the true interactive effects. At best, iFORM correctly detects an interaction as top five interactions in only 21\% of the replicates. Accompanied by the fact that both interactions are found to be important in just 9\% of the replicates, iFORM does not look like a promising approach for detecting interactions in this scenario. Note that JCI-SIS, on the other hand, accurately detects at least one of the two true interactions in every replication, and detects both true interactions in 84 of the 100 replications. 

\tfsib{
\subsection{Simulation Study 5}
Simulation 5 assesses the performance of JCI-SIS under various correlation parameters for continuous predictors when no main effects exist. We generate 100 samples for 500 predictors following a multivariate normal distribution with mean vector \textbf{0}. The correlations of $(X_1,X_2)$, $(X_3,X_4)$ and $(X_5, X_6)$ are 0.1, 0.3 and 0.5, respectively. We connect $Y$ with three pairs through 
\begin{center}
$Y= X_1 * X_2 + X_3 * X_4 + X_5 * X_6$.
\end{center}

\begin{table}[h!]
\caption{The percentage of replicates in the Simulation Study 5 for which the given true interaction(s) is found to be one of the five most important interactions
}\label{table:newSim}
\begin{center}
\begin{tabular}{ |c | c c | }
\hline
 & JCI-SIS & iFORM  \\ \hline\hline
$(X_1, X_2)$ & 78\% & 11\% \\
$(X_3, X_4)$ & 89\% & 15\% \\
$(X_5, X_6)$ & 96\% & 16\% \\
$(X_1, X_2), (X_3, X_4) \& (X_5, X_6)$ & 68\% & 4\% \\
  \hline
\end{tabular}
\end{center}
\end{table}

The results of the Simulation Study 5 are summarized in Table \ref{table:newSim}. As we can see that the accuracy of JCI-SIS in detecting the truly active pairs is at least six times higher than that of iFORM.
Comparing $(X_5, X_6)$ with $(X_1, X_2)$, we notice that the screening accuracy increases as the correlation gets stronger. This is to be expected because highly correlated predictors share more common information.} \tfsib{The run time is 464 seconds for JCI-SIS and 3300 seconds for iFORM for the 100 replications of the Simulation Study 5.}

\subsection{Real Data Analysis}
We apply \ta{the proposed JCI-SIS screening process} to a genome-wide association study \tfsib{downloaded from dbGaP} relating to polycystic ovary syndrome (PCOS) disease (dbGaP Study Accession:$https://www.ncbi.nlm.nih.gov/projects/gap/cgi-bin/study.cgi?study_id=phs000368.v1.p1).$ This data consists of 4099 (\ta{3056} controls and \ta{1043} cases) subjects, which are females who self-identified as having Caucasian or European-ancestry. The response is PCOS affection status (\tfsib{control/unaffected is coded as 1 and case/affected as 2}) and the predictors are the encoded SNP genotype values (1 for \emph{AA}, 2 for \emph{AB} or \emph{BA}, and 3 for \emph{BB}).  The modeling aim is to identify \ta{the most influential SNP interactions (epistasis) that are associated with} PCOS affection status.

\tfsib{The entire genome contains 731,442 raw SNPs. As \ta{a standard procedure}, the Plink software is utilized to perform quality control and remove those SNPs with minor allele frequency (MAF) $\le 5\%$, call rate $\ge 98\%$, $\ge 20\%$ missing genotype data, and p-value $\le 0.001$ for the Hardy-Weinberg Equilibrium (HWE) test \citep{anderson2009investigation, chen2011genome, hwang2012genome, purcell2007plink}. There are 595,935 SNPs and 4,098 people remaining after the quality control procedure.

There tends to be redundant information among regions of the genome in high linkage disequilibrium (LD), which harbor a set of SNPs that are inherited together. Therefore, we select a subset of tag SNPs as representatives within each LD block to reduce redundancy and greatly increase efficiency without the need of interrogating every SNP from the whole genomic dataset \citep{carlson2004selecting, barrett2005haploview,de2005efficiency}. The Haploview program written in JAVA is utilized to implement the Tagger algorithm based on the greedy pairwise LD tests using $r^2$ statistic. A threshold of 0.5 is set to select the maximally informative tag SNPs for further association analysis \citep{carlson2004selecting}. There are 234,754 representative SNPs selected from the Haploview program. 

JCI-SIS is then applied to all pairwise combinations of 234,754 SNPs across 23 \textit{homo sapien} chromosomes, which lead to 27,554,602,881 interaction pairs representing a very challenging ultrahigh dimensional setting. As a rough estimate, it would take 765,405 hours or 89.58 years using brute-force version of the original R code that costs around 0.1 second for each pair (the detailed computation is: $0.1\times 27,554,602,881/60/60/24/356=89.58$). 

To tackle this astronomic cost issue, we experiment the following multiple steps. 
\begin{itemize}
\item Step 1: Picking out all redundant and repeated computations from the doubled loops implementing Equation (\ref{Rhat}) to reduce the most straightforward overhead. After algorithm optimization, we identify that the terms, $\sqrt{n}$, $\bar{X}_{j1}$, $\bar{X}_{j2}$, $\sum_{i=1}^{n}\left ( X_{ij1} - \bar{X}_{j1} \right )^{2}$, $\sum_{i=1}^{n}\left ( X_{ij2} - \bar{X}_{j2} \right )^{2}$, and $\sum_{i=1}^{n}\left(Y_{i} - \bar{Y} \right)^{2}$, involve repeated computations. This step dramatically reduces the computational complexity ten thousand fold. The duration of each pair is reduced to 0.03 seconds with Python. 
\vspace{1em}

\item Step 2: As an attempt, we also transfer loops to matrices operation using NumPy with Python and then the computational time of each pair is compressed to 0.000052 seconds, which is around 1000 times faster than that of Step 1.  

\vspace{1em}

\item Step 3: To further alleviate computational suffer, we employ an open-source library CuPy for GPU-accelerated computing with Python because graphics processing unit (GPU) has been a powerful tool to exponentially increase speed. We take the NVIDIA GTX 1080ti as the co-processor to accelerate CPUs. However, even the most optimized GPU implementation is still not as fast as we expected. We speculate that the slight inefficiency of the GPU implementation is that some loops are still unavoidable and the matrix multiplication of this data analysis is not intensive enough. Moreover, it may consume too much time moving data from the CPU to the GPUs.    

\vspace{1em}

\item Step 4: Considering that most computationally-intensive R or Python packages are actually written in C++/C, we further evaluate our algorithm of Step 1 on a C++ version compiled with maximum optimization on Ubuntu. It leads to 0.0000036 seconds needed for one pair. 
\end{itemize}

Overall, the total computation time is dramatically reduced from 89.58 years years to 27.55 hours if using Step 1 + Step 4 on a single computer, as described above. To gain further computational speed-up, we parallelized nine independent cores of Intel i7-6850K to share the computational burden. All told, it takes only four hours to achieve the output for all 27,554,602,881 interaction pairs and complete the entire interaction screening process.

Figure \ref{f:Rhat} demonstrates the $\widehat{R}_{j_1, j_2}$ values for all 27,554,602,881 interaction pairs achieved by JCI-SIS  across the 23 chromosomes for this PCOS data analysis. The importance of all SNP pairs are ranked according to the  $\widehat{R}_{j_1, j_2}$ value. There are 96 interaction pairs having $\widehat{R}_{j_1, j_2}$ greater than 0.74. The top ten most important interaction pairs are summarized in Table \ref{table1}.

\begin{figure}[h!]
\hspace{-3em}
	\includegraphics[scale=0.4]{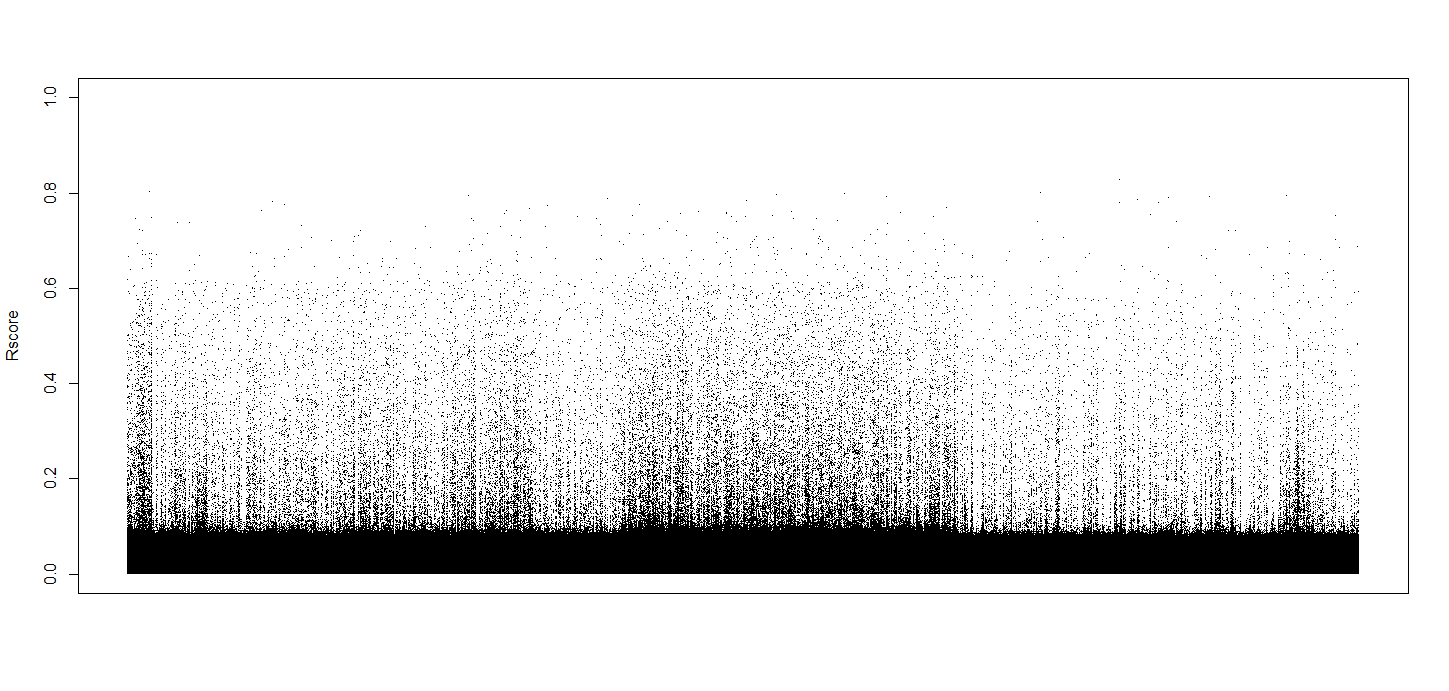}
	\vspace{-5em}
	\caption{The $\widehat{R}_{j_1, j_2}$ values from JCI-SIS for all 27,554,602,881 interaction pairs across the 23 chromosomes.}
	\label{f:Rhat}       
\end{figure}

\begin{table}[ht!]
\centering
\caption{\label{table1}The top ten most important two-way SNP interaction pairs for PCOS status.}
\begin{tabular}{|l|l|l|l|l|}
\hline
\textbf{SNP1}  & \textbf{SNP2} &  $\widehat{R}_{j_1, j_2}$ \\ \hline
ch13: rs4886241                       & ch13: rs728085               & 0.828114      \\ \hline
ch23: rs225066                   & ch23: rs225046                  & 0.802092      \\ \hline
ch12: rs10846902         & ch12: rs7956534                    & 0.799761      \\ \hline
ch10: rs4948591            & ch10: rs2047009                  & 0.798004      \\ \hline
ch11: rs3851173           & ch11: rs3907015                    & 0.794911      \\ \hline
ch18: rs12454453                  & ch18: rs12454349               & 0.7946        \\ \hline
ch6: rs1796520          & ch6: rs1407045                  & 0.793055      \\ \hline
ch2: rs2374292             & ch2: rs13016499           & 0.791583      \\ \hline
ch14: rs7159758               & ch14: rs17463975               & 0.790993      \\ \hline
ch13: rs9514046        & ch13: rs7993414                  & 0.789864      \\ \hline
\end{tabular}
\end{table}
The main objective is to discover the influential variables, rather than to measure their effects. Once they are detected, the problem of dealing with a much smaller group of influential variables can be yielded to appropriate analysis. 

Note that JCI-SIS can be used as a method that complements, rather than replaces, other in-depth interaction detecting approaches since the outputs of JCI-SIS can dramatically filter a large scale of noise two-way interactions and overcome the challenge of astronomic computational cost. Specifically, if higher order interactions is under interest, the multi-factor dimensionality reduction (MDR) has been found to be an effective model-free and nonparametric machine learning approach to identify interactions up to a certain order for case-control studies, even in the absence of \ta{individual main effects} \citep{RITCHIEmdr2001,Hahnmdr2003,Winham2011, GolaRoadmapMDR2016}. However, the maximum carrying capacity of the MDR is very limited. For example, it was claimed by \citet{Hahnmdr2003} that the software package implementing the MDR only allows for up to 15 genetic and/or environmental factors in a maximum of 400 study subjects and up to 500 total variables. Indeed, it is rare for in-depth approaches like MDR to be directly applicable for 27,554,602,881 interaction pairs.

As a follow up after the JC-SIS, we apply the MDR software written in JAVA to the top 96 interaction pairs having the largest values of $\Rhat$ ($\Rhat > 0.74$) to investigate the 3-way, 4-way, and 5-way ($d$ =3, 4, and 5) epistasis that influence PCOS, respectively \citep{PattinMDR2009,GreeneMDR2010}. The $d$ SNPs and all their possible multi-loci genotype combinations are represented in $d$-dimensional space, then the cases:controls ratio is estimated within each multi-loci cell. Because our case to control ratio is unbalanced (case:control ratio=0.34), we utilize the balanced accuracy (BA) that is constructed with an adjusted threshold $T_{adj}$ as the evaluation measure of prediction accuracy. The $T_{adj}$ \ta{can be set as} the ratio of cases:controls \citep{MDRunbalanced,GolaRoadmapMDR2016}, which is 0.34 in this dataset. The predicted PCOS status is then labeled as high risk (case:controls ratio > $T_{adj}$) or as low-risk (cases:controls ratio < $T_{adj}$) in each multi-loci cell in the $d$-dimensional space.

Finally, the average training errors and prediction errors across ten times 10-fold cross validation are reported to prevent overfitting and also avoid spurious results due to chance divisions of the data \citep{MDRunbalanced, Winham2011}. Single best multi-loci model with the minimum prediction error and the highest cross-validation consistency is selected for each of the three- to five-loci under consideration. The consistency is assessed as the number of cross-validation replications (maximum of 10) that a particular SNP combination is chosen by MDR across the 10 runs. 

Table \ref{MDR2} summarizes the cross-validation consistency and the average training and prediction balanced accuracies obtained from MDR for each of the 3-way, 4-way and 5-way interactions among the top 96 SNPs selected by JCI-SIS. We also implemented a single-SNP logistic regression model from PLINK and discovered no statistically significant evidence of independent main effects of any of these polymorphisms reported in Tables 7-8. Most of the p-values of these SNPs are very large, only one of them is less than 0.05 (0.0046 for chr1:rs267856).

\begin{table}[ht!]
\centering
\caption{\label{MDR2}Summary of results for higher-order interactions obtained from the MDR.}
\resizebox{\textwidth}{!}{
\begin{tabular}{|l|l|l|l|l|}
\hline
\textbf{Interaction}  & \textbf{Best Model}  &  \textbf{Training BA} & \textbf{Testing BA} \\ \hline
3-way & (chr1: rs267856, chr15: rs4414463, chr15: rs11634290)  & 0.777 & 0.766\\
4-way &  (chr10: rs4545483, chr10: rs3862015, chr15: rs4414463, chr15: rs11634290)  & 0.8881 & 0.8826\\
5-way & (chr11: rs12794303, chr11: rs11022159, chr15: rs4414463, chr15: rs11634290, chr21: rs465446)  & 0.8985 & 0.8702\\
 \hline
\end{tabular}
}
\end{table}

Figure \ref{f:threeway} summarizes the corresponding frequency of cases and of controls along with the predicted high risk (in dark grey) and low risk (in light grey), for each three-loci genotype cell. As an example, Figure \ref{f:threeway} visually demonstrates the epistasis. Specifically, the influence that each genotype at a particular locus (e.g., when rs267856=\emph{AB} and rs4414463=\emph{AB}) has on disease risk depends on the genotypes at each of the other loci (e.g., when rs11634290=\emph{AA, AB} or \emph{BB}). 

\begin{figure}[h!]
	\centering
	\includegraphics[scale=0.3]{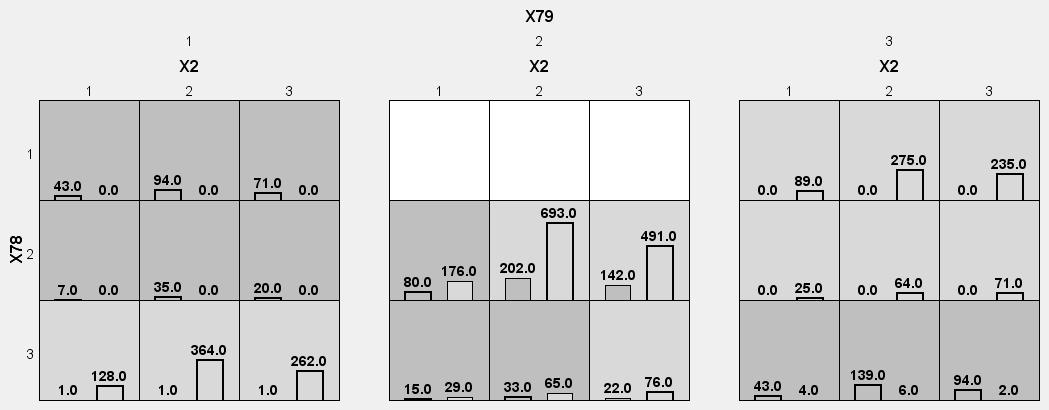}
	\caption{The frequency of cases and of controls along with the predicted high risk (in dark grey) and  low risk (in light grey), for each three-loci genotype cell. The empty cells are white.}
	\label{f:threeway}       
\end{figure}

There are around eight genes located from these informative SNP interaction pairs summarized in Tables 7-8. We confirm some genes that were already reported to be directly associated with ovary diseases or infertility. Specifically, the inter-chromosome 4-way interaction with the highest prediction balanced accuracy in Table 8 include the gene \emph{HABP2} (Chr10: rs4545483 and Chr10: rs3862015) and the gene \emph{CSPG4} (Chr15: rs4414463 and Chr15: rs11634290). The 3-way interaction also recognizes gene \emph{CSPG4} (chr15: rs4414463 and chr15: rs11634290). The \emph{CSPG4} has been found as an important target molecules in the top inhibited upstream regulators of miRNAs that involved in the complex ovarian follicular development and ovulation processes \citep{yerushalmi2018characterization} and also in female sterile lines with little or no ovulation or anovulatory infertility \citep{schmahl2008pdgf, liu2016network}. The contribution of the \emph{HABP2} gene to the risk of female infertility and in the regulation of endometrial angiogenesis was revealed by differentially expressed proteins identified in the follicular fluid from normal-weight patients with PCOS compared with the controls \citep{zhang2019tmt, altmae2009endometrial, altmae2011variation}. Moreover, the interaction of chr11: rs3851173 and chr11: rs3907015 in Table 7 both locate \emph{DLG2}. The gene \emph{DLG2} may contribute to delayed puberty and isolated hypogonadism \citep{jee2020dlg2} and also influence the differentially expressed long non-coding RNAs between the ovaries of multiparous and uniparous goat \citep{ling2017identification}. 

In addition to confirming the aforementioned three genes (\emph{DLG2, CSPG4}, and \emph{HABP2}) that were reported to be directly associated with ovary or infertility, we also detect some new genes: interaction of rs7159758 and rs17463975 in the gene \emph{NUBPL}; interaction of rs1796520 in the gene \emph{BTN3A1} and rs1407045 in the gene \emph{BTN2A1}; interaction of rs225066 and rs225046 in the gene \emph{IL1RAPL1}; and the interaction of rs10846902 and rs7956534 in the gene \emph{TMEM132B}. 

}

\section{DISCUSSION}\label{discuss}
\ta{The astronomic number of interaction terms to be considered for an ultrahigh dimensional setting represents unprecedented challenges. The majority of classic approaches designed for modeling interaction become either computationally infeasible or lose accuracy and robustness as the dimension is extremely large. Several powerful feature screening approaches have been developed to address the challenges in ultrahigh dimensional data, however interaction screening has received little attention compared to individual main effects. A general} issue that limits the capability of existing interaction feature screening approaches is the over-reliance on the existence of pronounced main effects.

\ta{In this paper we proposed a novel interaction feature screening approach, JCI-SIS, that is empirically accurate, theoretically sound, and computationally feasible for ultrahigh dimensional feature spaces. The advantages of JCI-SIS are its versatility and practicability to extract interactive effects regardless of data type or the existence of main effects. We prove that the JCI-SIS satisfies the strong sure screening consistency property, meaning that JCI-SIS prevails in discovering the exact set of \ta{truly influential features} with probability approaching one as the \ta{number of features goes to infinity. The subset of this property, sure screening consistency,} has become the benchmark theoretical property \ta{for judging valid} feature screening methods. 

\ta{The capability of JCI-SIS was demonstrated through multiple finite sample simulation studies comparing with the most relevant approaches in the literature. JCI-SIS improved in both speed and accuracy} when compared to other existing interaction screening methods such as iFORM \citep{HaoZhangiFORM} and GPC-SIS \citep{Huang2014}.} The first two simulation designs lead us to believe that, in the absence of main effects, JCI-SIS produces competent results without relying on a heredity structure. Moreover, the last two simulation designs show that even when strong main effects are present (a scenario that favors iFORM and GPC-SIS), JCI-SIS \ta{still} excels in determining the truly influential interactions. Under empirical observation, JCI-SIS ran about six to seven times faster than GPC-SIS when applied to the same data sets on the same machine. 

We furthermore gain confidence in the applicability of JCI-SIS to extremely large real life data sets. (\tfsib{Our real data analysis processed about 27,554,602,881 pairwise interactions}). In terms of number of interaction pairs in question, as well as the number of observations considered, the real data set we examine is approximately \tfsib{ten thousand} times larger than the data sets examined in similar papers (e.g. the inbred mouse microarray gene expression dataset found in \cite{HaoZhangiFORM}). In turn, this means that the computational considerations necessary for our real data analysis were, until now, unseen in the setting of interaction feature screening.

\section{APPENDIX: Proofs of Theoretical Results}\label{proofSection}

Here we present in full the proofs for Theorems 1 and 2. The following lemma will assist in the proof of our main theorems.

\ta{In the context of the denominator of the definition of $\widehat{R}_{j_1, j_2}$, we define
\[
\hat{\sigma}_{j_1}=\sqrt{\frac{1}{n}\left(\sum\limits_{i = 1}^{n}(X_{ij_1} - \overline{X}_{j_1})^2\right)}
\]
as an estimator of $\sqrt{\kappa_2(X_{j_1}, X_{j_1})}$, 
\[
\hat{\sigma}_{j_2}=\sqrt{\frac{1}{n}\left(\sum\limits_{i = 1}^{n}(X_{ij_2} - \overline{X}_{j_2})^2\right)}
\]
as an estimator of $\sqrt{\kappa_2(X_{j_2},X_{j_2})}$. Similarly, we define
\[
\hat{\sigma}_Y=\sqrt{\frac{1}{n}\left(\sum\limits_{i = 1}^n(Y_i-\overline{Y})^2\right)}
\]
as an estimator of $\sqrt{\kappa_2(Y,Y)}$. In the context of the numerator of $\widehat{R}_{j_1, j_2}$, we define
\begin{equation}\label{tauhat}
\hat{\tau}_{j_1, j_2}=\frac{1}{n}\sum\limits_{i = 1}^n(X_{ij_1} - \overline{X}_{j_1})(X_{ij_2} - \overline{X}_{j_2})(Y_i - \overline{Y})
\end{equation}
as an estimator of $\kappa_3(Y, X_{j_1}, X_{j_2})$.}

\vspace{2em}

\textbf{Lemma A.1.} $\widehat{R}_{j_1, j_2}$ is a consistent estimator of $R_{j_1, j_2}$.
\ta{As a well established result, we know that $\hat{\sigma}^2_{j_1}$, $\hat{\sigma}^2_{j_2}$, and $\hat{\sigma}^2_Y$ are all (individually speaking) bias yet \textit{consistent} estimators of their respective population variance.} We now apply the weak law of large numbers to show that $\hat{\tau}_{j_1, j_2}$ defined in Equation (\ref{tauhat}) is a consistent estimator of $\kappa_3(Y, X_{j_1}, X_{j_2})$.

\ta{Explicitly expanding the product of binomials in (\ref{tauhat}) we  get}
\begin{eqnarray*}
\hat{\tau}_{j_1, j_2} &=& \frac{1}{n}\sum X_{ij_1}X_{ij_2}Y_i - \frac{1}{n}\sum \overline{X}_{j_1}X_{ij_2}Y_i - \frac{1}{n}\sum X_{ij_1}\overline{X}_{j_2}Y_i \\\\
&&\quad\quad\quad - \frac{1}{n}\sum X_{ij_1}X_{ij_2}\overline{Y} + \frac{1}{n}\sum \overline{X}_{j_1}\overline{X}_{j_2}Y_i\\\\
&&\quad\quad\quad + \frac{1}{n}\sum \overline{X}_{j_1}{X}_{ij_2}\overline{Y}+ \frac{1}{n}\sum {X}_{ij_1}\overline{X}_{j_2}\overline{Y}-\frac{1}{n}\sum \overline{X}_{j_1}\overline{X}_{j_2}\overline{Y}\\\\
&\xrightarrow{p}& \EE(YX_{j_1}X_{j_2}) - \EE(YX_{j_1})\EE X_{j_2}-\EE(YX_{j_2})\EE X_{j_1}\\\\
&& \quad \quad \quad \quad \quad ~-\EE(X_{j_1}X_{j_2})\EE Y + 2 \EE Y\EE X_{j_1}\EE X_{j_2}\\\\
&=& \kappa_3(X_{j_1}, X_{j_2}, Y).
\end{eqnarray*}
\ta{Converge in probability is guaranteed here by repeated applications (summand wise) of the weak law of large numbers to each term of $\hat{\tau}_{j_1, j_2}.$ That is}, $\frac{1}{n}\sum X_{ij_1}X_{ij_2}Y_i \xrightarrow{p} \mathbb{E}(X_{j_1}X_{j_2}Y)$, $\frac{1}{n}\sum \overline{X}_{j_1}X_{ij_2}Y_i \xrightarrow{p} \mathbb{E}(X_{j_1})\mathbb{E}(X_{j_2}Y)$, and \ta{$\frac{1}{n}\sum {X}_{ij_1}\overline{X}_{j_2}\overline{Y} \xrightarrow{p}  \mathbb{E}(X_{j_1}) \mathbb{E}(X_{j_2})\mathbb{E}(Y)$} and similar conclusions can be reached for other terms. \ta{This complete the proof about} $\hat{\tau}_{j_1, j_2}$ is a consistent estimator of $\kappa_3(X_{j_1}, X_{j_2}, Y)$. 

Since \ta{$\widehat{R}_{j_1, j_2}= |\hat{\tau}_{j_1, j_2}|/(\hat{\sigma}_{j_1}\hat{\sigma}_{j_2}\hat{\sigma}_Y)$} is a continuous function involving $\hat{\sigma}_{j_1}$, $\hat{\sigma}_{j_2}$, $\hat{\sigma}_Y$, and $\hat{\tau}_{j_1, j_2}$, a direct application of the Continuous Mapping Theorem \ta{or Mann-Wald Theorem} yields that $\widehat{R}_{j_1, j_2}$ is a consistent estimator of $R_{j_1, j_2} $ \citep{MannWald1943, Casella2002statistical}.



The proof of the two theorems is accomplished in three steps:
\begin{enumerate}
    \item We first show that a positive lower bound $R_{\min}$ exists for all $R_{j_1, j_2}$ with $(j_1, j_2) \in \mathcal{S}_T$. In other words, we will show the following:
     \[\text{There exists}~~ R_{\min} >0 \text{ such that } R_{j_1, j_2} > R_{\min}~~ \text{for all} ~~(j_1, j_2) \in \mathcal{S}_T.\]

    \item This is followed by our showing that $\Rhat$ is a uniformly consistent estimator of $R_{j_1, j_2}$ for each $1 \leq j_1 < j_2 \leq p$. 

    \item We finally show that there exists a constant $c > 0$ such that
    \[\mathbb{P}(\widehat{\mathcal{S}} = \mathcal{S}_T) \longrightarrow 1 \text{ as } n \longrightarrow \infty.\]
    Weak consistency is shown as a natural subcase of this, which will establish Theorem 2.1.
\end{enumerate}

\subsubsection*{Step 1}
\ta{
As defined in the main article, 
\[R_{j_1, j_2} = \frac{|\kappa_3(Y, X_{j_1}, X_{j_2})|}{\sqrt{\kappa_2(X_{j_1},X_{j_1} )\kappa_2(X_{j_2}, X_{j_2})\kappa_2(Y,Y)}},~~j_1<j_2;~j_1,j_2=1,\ldots,p.\]
For any pair of indices $(j_1, j_2) \in \mathcal{S}_T,$
\begin{eqnarray*}
R_{j_1, j_2} &\geq& \frac{|\kappa_3(Y, X_{j_1}, X_{j_2})|}{\sigma_{\text{max}}^3} \quad \text{by (C2),}\\[1.25ex]
&\geq& \frac{\delta_{\min}}{\sigma_{\text{max}}^3} \quad \text{by (C3),}\\[1.25ex]
&>& 0.
\end{eqnarray*}
}

Define $R_{\min} = \frac{\delta_{\min}}{2\sigma_{\text{max}}^3}.$
Then $R_{j_1, j_2} > R_{\min} > 0$ for all $(j_1, j_2) \in \mathcal{S_T}$.
\vspace{0.3cm}
\ta{This establishes Step 1 of the proof (i.e., Corollary 1).}

\subsubsection*{Step 2}\label{Step2}
We will now show that such consistency is also uniform. 
Since $\Rhat$ is consistent as an estimator of $R_{j_1, j_2}$ (\ta{see Lemma A.1}), we know that for any $1 \leq j_1 < j_2 \leq p$ and any  $\varepsilon > 0$,
\[\mathbb{P}(|\Rhat - R_{j_1, j_2}| > \varepsilon) \rightarrow 0 \quad \text{ as } n\rightarrow\infty.\]
Let \[(J_1, J_2) = \text{argmax}_{1 \leq j_1 < j_2 \leq p}~|\Rhat - R_{j_1, j_2}|.\]
Then, we have that
\[\mathbb{P}(|\widehat{R}_{J_1, J_2} - R_{J_1, J_2}| > \varepsilon) \rightarrow 0 \quad \text{ as } n\rightarrow\infty\] for any $\varepsilon> 0$. In other words, we have that
\[\mathbb{P}\left(\max_{1 \leq j_1 < j_2 \leq p}|\Rhat - R_{j_1, j_2}| > \varepsilon\right) \rightarrow 0 \quad \text{ as } n\rightarrow\infty\] for any $\varepsilon> 0$.
This shows that $\Rhat$ is a \textit{uniformly} consistent estimator of $R_{j_1, j_2}$, completing Step 2. This also establishes Corollary 2.

\subsubsection*{Step 3}
Let $c = (2/3)R_{\min}$. Suppose by way of contradiction that this cutoff is insufficient to be able to claim \ta{$\mathcal{S}_T  \subseteq \widehat{\mathcal{S}}$}. This would mean that there exists some pair $(j_1^*, j_2^*) \in \mathcal{S}_T$, yet $(j_1^*, j_2^*) \notin \widehat{\mathcal{S}}$. It then follows that we must have \[\widehat{R}_{j_1^*, j_2^*} \leq c=(2/3)R_{\min}\] \ta{by the definition of $\widehat{\mathcal{S}}$}, while at the same time having (as shown in Step 1) \[R_{j_1^*, j_2^*}> R_{\min}.\]

From this we can conclude that \[|\widehat{R}_{j_1^*, j_2^*} - R_{j_1^*, j_2^*}| > (1/3)R_{\min},\] which implies that \[\max_{1 \leq j_1< j_2 \leq p}~|\Rhat - R_{j_1, j_2}| > (1/3)R_{\min}\] as well. However, we know by the uniform consistency of $\Rhat$ that by letting $\varepsilon = 1/3 R_{\min}$, we have
\[\mathbb{P}(\ta{\mathcal{S}_T  \not \subseteq \widehat{\mathcal{S}}}) \leq \mathbb{P}\left(\max_{1 \leq j_1 < j_2 \leq p}|\Rhat - R_{j_1, j_2}| >(1/3)R_{\min}\right) \rightarrow 0 \quad \text{ as } n\rightarrow\infty.\] This is a contradiction to the assumption of non containment above. So indeed, we have that \[\mathbb{P}(\ta{\mathcal{S}_T  \subseteq \widehat{\mathcal{S}}})\rightarrow 1\quad \text{ as } n\rightarrow\infty.\]
This proves Theorem 1, and also establishes the forward direction for the statement of Theorem 2.

To prove the reverse direction for Theorem 2, suppose (again by way of contradiction) that $\widehat{\mathcal{S}} \not\subseteq \mathcal{S}_T$. Then there is some $(j_1^*, j_2^*)\in \widehat{\mathcal{S}}$, yet $(j_1^*, j_2^*) \notin \mathcal{S}_T.$ This means that (\ta{by the definition of $\widehat{\mathcal{S}}$}) \[\widehat{R}_{j_1^*, j_2^*} \geq (2/3)R_{min},\] while at the same time (by (C4)) having \[R_{j_1^*, j_2^*} = 0.\] It now follows that \[|\widehat{R}_{j_1^*, j_2^*} - R_{j_1^*, j_2^*}| > (2/3)R_{\min}.\] Set $\varepsilon = (2/3)R_{\min}$. By uniform consistency we have
\[\mathbb{P}(\ta{\widehat{\mathcal{S}} \not\subseteq \mathcal{S}_T}) \leq \mathbb{P}\left(\max_{1 \leq j_1< j_2 \leq p}|\Rhat - R_{j_1, j_2}| >(2/3)R_{\min}\right) \rightarrow 0 \quad \text{ as } n\rightarrow\infty.\]
From this we know that
\[\mathbb{P}(\ta{\widehat{\mathcal{S}} \subseteq \mathcal{S}_T})\rightarrow 1\quad \text{ as } n\rightarrow\infty.\]
We can now conclude that for $c = (2/3)R_{\min}$, we have $\mathbb{P}(\mathcal{S}_T = \widehat{\mathcal{S}}) \rightarrow 1$ as $ n\rightarrow \infty$, completing the proof.

\section{DISCLOSURE DECLARATIONS}

\subsection{Conflicts of Interest}
The authors declare that there is no conflict of interest.

\subsection{Availability of Data and Material}
The dataset(s) used for the analyses described in this manuscript were obtained from the
database of Genotype and Phenotype (dbGaP) found at http://www.ncbi.nlm.nih.gov/gap
through dbGaP accession number phs000368. Samples and associated phenotype data for the
Genome-Wide Association Scan [GWAS] of Polycystic Ovary Syndrome Phenotypes were provided
by Andrea Dunaif, M.D.

\subsection{Code Availability}
The computer code is available upon request.



\bibliographystyle{spbasic}      
\bibliography{BibliographyScreening}

\end{document}